\documentclass[10pt]{sn-jnl}% Vancouver Reference Style
\usepackage{graphicx}%
\usepackage{multirow}%
\usepackage{amsmath,amssymb,amsfonts}%
\usepackage{amsthm}%
\usepackage{mathrsfs}%
\usepackage[title]{appendix}%
\usepackage{xcolor}%
\usepackage{textcomp}%
\usepackage{manyfoot}%
\usepackage{booktabs}%
\usepackage{algorithm}%
\usepackage{algorithmicx}%
\usepackage{algpseudocode}%
\usepackage{listings}%
\theoremstyle{thmstyleone}%
\theoremstyle{thmstyletwo}%

\theoremstyle{thmstylethree}%

\begin{document}

\title[Article Title]{Elasticity affects the shock-induced aerobreakup of a polymeric droplet}

%%=============================================================%%
%% GivenName	-> \fnm{Joergen W.}
%% Particle	-> \spfx{van der} -> surname prefix
%% FamilyName	-> \sur{Ploeg}
%% Suffix	-> \sfx{IV}
%% \author*[1,2]{\fnm{Joergen W.} \spfx{van der} \sur{Ploeg} 
%%  \sfx{IV}}\email{iauthor@gmail.com}
%%=============================================================%%

\author[1]{\fnm{Navin Kumar} \sur{Chandra}}\email{navinchandra@iisc.ac.in}

\author[1,2]{\fnm{Shubham} \sur{Sharma}}\email{ssharm94@jh.edu}

\author[1,3]{\fnm{Saptarshi} \sur{Basu}}\email{sbasu@iisc.ac.in}

\author*[1]{\fnm{Aloke} \sur{Kumar}}\email{alokekumar@iisc.ac.in}

\affil*[1]{\orgdiv{Department of Mechanical Engineering}, \orgname{Indian Institute of Science Bangalore}, \orgaddress{\postcode{560012}, \state{Karnataka}, \country{India}}}

\affil[2]{\orgdiv{Department of Mechanical Engineering}, \orgname{Johns Hopkins University, Baltimore}, \orgaddress{\postcode{21218}, \state{MD}, \country{USA}}}

\affil[3]{\orgdiv{Interdisciplinary Centre for Energy Research}, \orgname{Indian Institute of Science Bangalore}, \orgaddress{\postcode{560012}, \state{Karnataka}, \country{India}}}

%%==================================%%
%% Sample for unstructured abstract %%
%%==================================%%

\abstract{Boger fluids are viscoelastic liquids having constant viscosity for a broad range of shear rates. They are commonly used to separate the effects of liquid elasticity from viscosity in any experiment. We present an experimental study on the shock-induced aerobreakup of a Boger fluid droplet in the Shear-induced entrainment (SIE) and catastrophic breakup regime (Weber number ranging from $\sim$ 800 to 5000). The results are compared with the aerobreakup of a Newtonian droplet having similar viscosity, and with shear-thinning droplets. The study aims to identify the role of liquid elasticity without the added complexity of simultaneous shear-thinning behavior. It is observed that at the early stages of droplet breakup, liquid elasticity plays an insignificant role, and all the fluids show similar behavior. However, during the late stages, the impact of liquid elasticity becomes dominant, which results in a markedly different morphology of the fragmenting liquid mass compared to a Newtonian droplet.}


%%\pacs[JEL Classification]{D8, H51}

%%\pacs[MSC Classification]{35A01, 65L10, 65L12, 65L20, 65L70}

\maketitle

\section{Introduction}\label{intro}
In the context of secondary atomization, the word aerobreakup refers to the process of breaking a liquid droplet into smaller fragments by subjecting it to a high-speed stream of gas (generally air). Understanding the physics of aerobreakup lies at the core of many natural and industrial processes like fuel atomization inside an internal combustion engine, the breakup of sneezed ejecta \cite{scharfman2016visualization}, spray atomization of industrial chemicals \cite{qian2021experimental}, liquid metal atomization \cite{Sharma_Chandra_Kumar_Basu_2023} and the breakup of a falling raindrop \cite{villermaux2009single}, etc. Compared to non-Newtonian droplets, the aerobreakup of a Newtonian droplet is a relatively well-studied problem, and a comprehensive understanding can be gained from various review articles \cite{PILCH_1987, gelfand1996droplet, guildenbecher2009secondary, theofanous2011aerobreakup, sharma2022advances}. However, the research in the aerobreakup of non-Newtonian droplets has received less attention, despite their present and potential future applications in many areas. The research in this area is so sparse that there is a lack of common consensus on something as basic as the suitable non-dimensional number that governs the breakup process \cite{guildenbecher2009secondary}. There are a few early studies on the shock-induced aerobreakup of viscoelastic droplets \cite{wilcox1961retardation, matta1982viscoelastic, matta1983aerodynamic, arcoumanis1994breakup}. While all these studies agree on the conclusion that the presence of liquid elasticity hinders aerobreakup, the precise mechanism through which it becomes operational remains elusive. Joseph et al. \cite{joseph2002rayleigh} suggested that the catastrophic breakup at a very high Weber number ($\sim10^4-10^5$) happens through the mechanism of Rayleigh-Taylor instability. However, their theory predicted that the instability growth rate for a viscoelastic fluid is higher than that of a Newtonian fluid with similar viscosity, which contrasts with the widely accepted experimental observations \cite{chandra2023shock, arcoumanis1994breakup, matta1982viscoelastic}. The problem of the non-Newtonian droplet aerobreakup becomes more challenging because non-Newtonian behavior in a liquid can arise in various ways, such as yield stress, shear rate-dependent viscosity (thinning or thickening), viscoelasticity, and a combination of these behaviors. In our previous works \cite{chandra2023shock, chandra2023aerodynamic}, we delved into the mechanism by which liquid elasticity influences the aerobreakup of a polymeric droplet. Our findings indicate that liquid elasticity has a negligible impact during the initial stages of the breakup phenomenon, where droplet deformation and the growth of hydrodynamic instabilities (Kelvin-Helmholtz and Rayleigh-Taylor instability) occur. This lack of influence is attributed to the insufficient strain and strain rate during the early stage, preventing significant stretching of polymer molecules and maintaining behavior akin to Newtonian fluids. However, as the breakup progresses to its later stages, large deformations at sufficiently high deformation rates in the liquid phase result in elastic stresses becoming prominent due to polymer stretching. These elastic stresses provide significant resistance against the fragmentation of the liquid mass, altering the final morphology compared to a Newtonian liquid. It should be noted that in these previous studies, the viscosity of the polymeric solutions differed from that of the Newtonian solvent, and some of these polymeric solutions displayed shear-thinning behavior. Therefore, it remains an unresolved issue to isolate the effect of elasticity from the viscosity and shear-thinning behavior.

To resolve the above-mentioned issue, we employ Boger fluids, which do not show a strain rate-dependent viscosity in shear flows (like a Newtonian liquid) but exhibit viscoelastic properties \cite{james2009boger}. By comparing the aerobreakup of a Boger fluid droplet with that of a Newtonian droplet having similar shear viscosity, we get rid of any effect that may come from the shear-thinning behavior. We also compare the present results with that of the shear-thinning droplets from our previous work \cite{chandra2023shock} to show that our previous conclusions on the influence of elasticity in modulating aerobreakup are valid even without shear-thinning behavior.

\section{Experimental setup}\label{exp_setup}
\begin{figure}[ht]%
\centering
\includegraphics[width=0.8\textwidth]{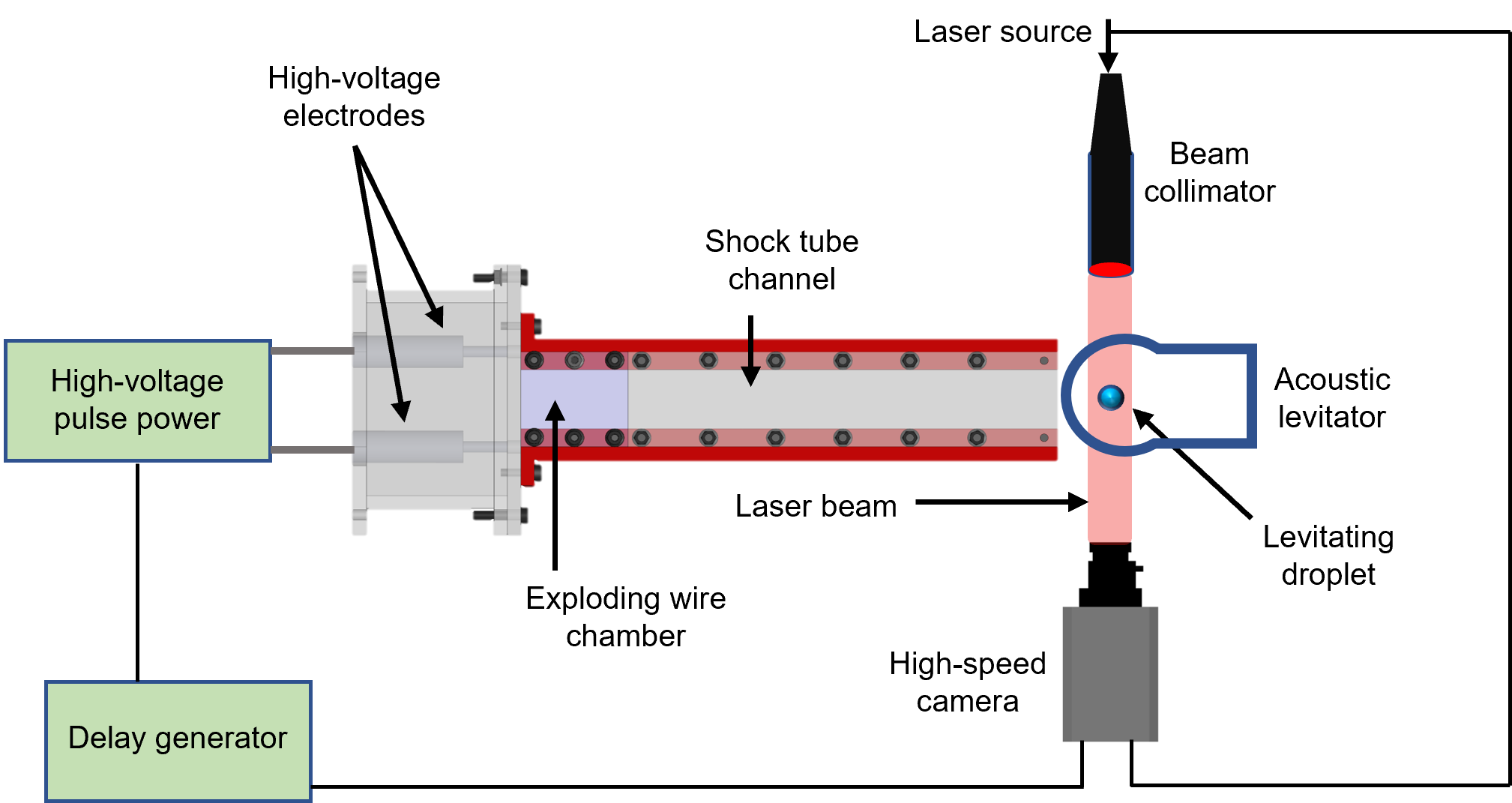}
\caption{Schematic diagram of the experimental setup.}
\label{fig:exp_setup}
\end{figure}

Experiments are performed in an exploding-wire-based shock tube setup, as shown in Figure \ref{fig:exp_setup}. This is the same setup we used in our previous work, and the detailed characteristics of this setup can be found there \cite{chandra2023shock, sharma2021shock, sharma2023depth}. In this setup, the sudden discharge of electrical energy from a high-voltage pulse power system (Zeonics Systech India, Z/46/12) leads to the explosion of a thin copper wire and the subsequent formation of a blast wave. This moving blast wave is focused and transformed into a normal shock using a narrow shock tube channel (320 mm $\times$ 50 mm $\times$ 20 mm). An acoustically levitated liquid droplet is kept stationary at the exit of the shock tube channel. This liquid droplet undergoes the aerobreakup process by interacting with the strong airflow induced behind the moving blast wave. A high-speed camera (Photron SA5) coupled with an ultra-high-speed pulsed laser light source (Cavitar Cavilux smart UHS), as shown in Figure \ref{fig:exp_setup}, is employed to record the shadowgraphic images of the droplet breakup phenomenon at a recording rate of 40000 frames per second. To get a global view of the breakup phenomenon, a macro lens (Sigma DG 105 mm) is connected to the high-speed camera, and recordings are made with a spatial resolution of $\approx$ 41 $\mu$m/pixel, having a field of view as $\approx$ 26 mm $\times$ 11 mm. A precise synchronization between the blast wave generation and the high-speed camera triggering is achieved using a delay generator (BNC 575).

\section{Materials and characterization}\label{materials}
Boger fluids are prepared by following the method suggested by Dontula et al. \cite{dontula1998model}. It is an aqueous solution of two different polymers mixed together. A low molecular weight ($\approx10^4$ g/mol) polyethylene glycol (PEG) is used to increase the shear viscosity of the solution (without imparting shear-thinning nature), while a minute concentration of high molecular weight ($\approx5\times10^5$ g/mol) polyethylene oxide (PEO) is used to impart viscoelastic properties. A low viscosity Boger fluid with shear viscosity $\approx$ 20 mPa-s, referred to as B1 consists of PEG and PEO with respective weight fractions as 20\% and 0.05\% in DI water. The high viscosity Boger fluid with shear viscosity $\approx$ 100 mPa-s, referred as B2, consists of 40\% PEG and 0.15\% PEO in DI water. The Newtonian counterparts of these Boger fluids, referred to as N1 and N2, are prepared by varying the glycerol content in a water-glycerol mixture to approximately match the shear viscosity of B1 and B2 fluids. The weight fraction of glycerol in N1 and N2 fluids are 70\% and 86\%, respectively. The data for shear-thinning fluids is taken from our previous work \cite{chandra2023shock}. These shear-thinning fluids, referred to as ST1 and ST2 in the present work, are aqueous solutions of PEO with concentrations of 0.4\% and 1\%, respectively. The physical properties and composition of all the test liquids are summarized in Table \ref{table_1}.
\begin{table}[h]
\caption{Fluid properties}\label{table_1}%
\begin{tabular}{@{}llllllll@{}}
\toprule
Name & Composition  & $\rho_l$ & $\gamma$ & $\mu_l$ & $\lambda$ & $Oh$ & $El$ \\
 & (\% w/w) & (kg/m$^3$) & (mN/m) & (mPa-s) & (ms) &  & \\
\midrule
N1 & Glycerol-70 & 1178 & 65 & 20 & - & 0.05 & -  \\
\medskip
 & Water-30 &  &  &  &  &  &  \\

N2 & Glycerol-86 & 1224 & 65 & 100 & - & 0.23 & -  \\
\medskip
 & Water-14 &  &  &  &  &  &  \\

B1 & PEG-20 & 1026 & 62 & 20 & 8.9 & 0.06 & 0.04  \\
\medskip
 & PEO-0.05 &  &  &  &  &  &  \\

B2 & PEG-40 & 1056 & 62 & 100 & 178 & 0.28 & 4.21  \\
\medskip
 & PEO-0.15 &  &  &  &  &  &  \\

\medskip
ST1 & PEO-0.4 & 1000 & 62 & 77\footnotemark[1] & 46 & 0.22 & 0.89  \\

\medskip
ST2 & PEO-1 & 1000 & 62 & 2081\footnotemark[1] & 134 & 5.91 & 69.7  \\

\botrule
\end{tabular}
\footnotetext{Note: In the composition column, polymer concentrations are reported as the weight percent of the solvent (water). $Oh$ and $El$ values are calculated considering a droplet diameter of 2 mm.}
\footnotetext[1]{Zero-shear viscosity}
\end{table}

\begin{figure}[ht]%
\centering
\includegraphics[width=0.95\textwidth]{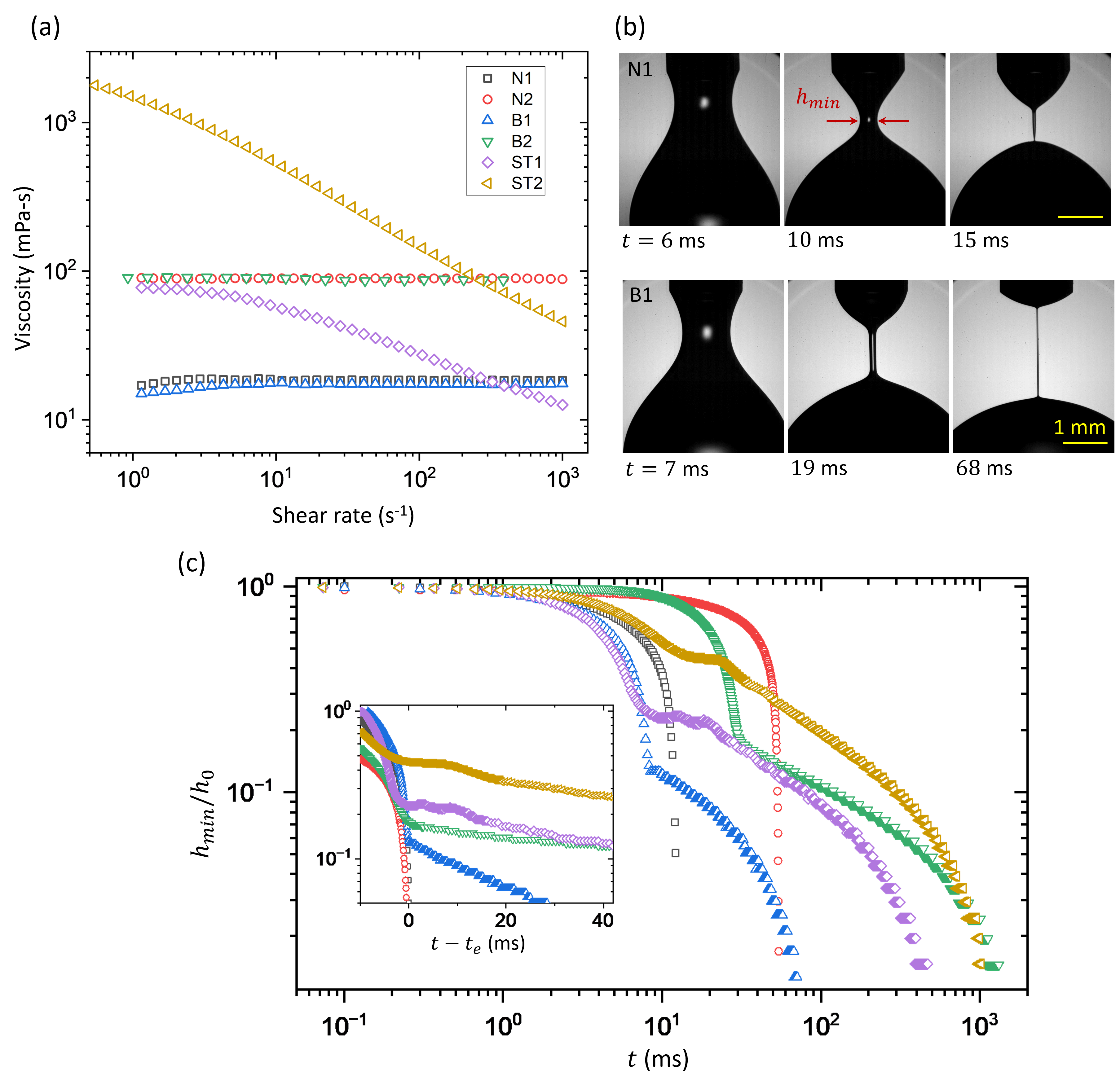}
\caption{Rheological characterization of the test liquids. (a) Shear viscosity at different shear rates obtained using a commercial rheometer. (b) Experimental snapshots for N1 and B1 fluids, obtained from the custom-built DoS rheometry setup for probing the extensional rheology. (c) Transient evolution of minimum neck diameter for all the liquids obtained from DoS rheometry experiments. The color and shape of symbols used in (c) represent the same fluids as in (a).}
\label{fig:rheology}
\end{figure}

The density, $\rho_l$, of the test liquids is estimated by weighing a known volume of the liquid. Surface tension, $\gamma$ is measured using an optical contact angle measuring and contour analysis system (OCA25) instrument from Dataphysics by the pendant drop method. The shear viscosity of test liquids ($\mu_l$), shown in Figure \ref{fig:rheology}a, are obtained using a cone and plate geometry CP-40 (plate diameter=40 mm, cone angle=1$^{\circ}$) of a commercial rheometer (Anton Paar, MCR302). It can be observed that Boger fluids (B1, B2) do not show shear-rate dependent viscosity variation, similar to that of the Newtonian liquids (N1, N2). However, the shear-thinning fluids (ST1, ST2) show approximately one order decrease in viscosity when the shear rate is varied from $\sim$1 to 1000 s$^{-1}$. The elastic properties of the test liquids are probed in extensional flows using a custom-built dripping-onto-substrate (DoS) rheometry setup following the standard protocol mentioned in literature \cite{dinic2015extensional, dinic2017pinch}. In DoS rheometry, an unstable liquid bridge is formed between a needle and a substrate by depositing a millimeter-sized droplet from the needle onto the substrate. The liquid bridge pinches off under the action of capillary force, and the time evolution of minimum neck diameter, $h_{min}$ of the liquid bridge is recorded. Experimental snapshots of typical DoS experiments are shown in Figure \ref{fig:rheology}b for N1 and B1 fluids. Figure \ref{fig:rheology}c shows the transient evolution of $h_{min}$ for all the test liquids such that $t=0$ corresponds to the time instant of $h_{min}/h_0=1$ with $h_0$=1.65 mm being the outer diameter of the needle used for DoS experiments. In the case of Newtonian liquids, $h_{min}$ follows a power law relation with time. However, in the case of viscoelastic liquids, the time evolution of $h_{min}$ is initially similar to the Newtonian liquid, but it shows a transition to the exponentially decaying regime at a later stage, known as the elasto-capillary regime \cite{dinic2015extensional}. This regime appears as a straight line in a semi-log plot with shifted time on the abscissa, as shown in the inset of Figure \ref{fig:rheology}c such that $t_e$ represents the time instant of transition from Newtonian to the elasto-capillary regime. The extensional relaxation time, $\lambda$ of a viscoelastic liquid, is obtained by fitting an exponential curve, $h_{min}=Ae^{-B(t-t_e)}$ in the elasto-capillary regime such that $\lambda=\frac{1}{3B}$. Average values of $\lambda$ for the present viscoelastic liquids are presented in Table \ref{table_1}.

\section{Breakup modes and suitable dimensionless groups}\label{modes}
There is abundant literature on the aerobreakup of a Newtonian droplet \cite{guildenbecher2009secondary}, and it is well established that the two most important non-dimensional numbers governing the aerobreakup process are Weber number, $We$, and Ohnesorge number, $Oh$ defined as follows-

\begin{equation}
We = \frac{\rho_g U_g^2 D_0}{\gamma}
\label{eq1}
\end{equation}

\begin{equation}
Oh = \frac{\mu_l}{\sqrt{\rho_l \gamma D_0}}
\label{eq2}
\end{equation}

Here, $\rho_g$ and $U_g$ are the density and free stream velocity of the gas phase. $D_0$ is the initial diameter of the droplet. Classically, breakup modes are categorized on a $We-Oh$  number plane and identified based on the morphology of the liquid mass \cite{HSIANG_1992}. For low-viscosity liquids ($Oh<$0.1), the breakup modes are independent of $Oh$ and only governed by the $We$ value. These modes, with their increasing order of $We$ range are- vibrational ($We<11$), bag ($11<We<18$), bag-stamen ($18<We<35$), multi-bag ($35<We<80$), shear stripping ($80<We<350$), and catastrophic breakup mode ($We>350$) \cite{guildenbecher2009secondary, jackiw_ashgriz_2021, HSIANG_1992}. The exact value of $We$ slightly varies among different literature. Later, Theofanous et al. \cite{theofanous2008physics, theofanous2011aerobreakup, theofanous2012physics, theofanous2013physics} categorized the breakup modes into two broad categories- Rayleigh-Taylor piercing (RTP) and shear-induced entrainment (SIE) mode, based on the governing hydrodynamic instabilities. The former is observed at low $We$($<$100) and governed by the Rayleigh-Taylor (RT) instability, while the latter is observed at high $We$($>$1000) and governed by the Kelvin-Helmholtz (KH) instability. The intermediate regime (100$<We<$1000) is identified as the transition regime between the two modes.

All the breakup modes discussed above primarily come from the experiments performed either using a continuous airflow through a nozzle or a traditional diaphragm-based shock tube setup. In these kinds of setup, the flow properties of the gas phase remain constant in the timescales of droplet breakup. However, the present setup uses a blast wave, which generates an airflow that decays continuously with time. Therefore, the breakup modes and their corresponding $We$ range obtained from the present setup can not be directly compared with the existing literature. It should be noted that the Weber number reported in the present work is based on the airflow velocity at the time of droplet-shock wave interaction. A detailed discussion of the airflow characteristics of the present setup is provided in our previous work \cite{chandra2023shock}. We have identified three different breakup modes obtained from the present setup. These modes are- vibrational ($700<We$), SIE ($700<We<2800$), and catastrophic ($We>2800$). In the vibrational regime, the aerodynamic forces are insufficient to cause significant liquid mass breakup. Therefore, the present work focuses mainly on the SIE regime and the subsequent transition to the catastrophic regime.

\begin{figure}[ht]%
\centering
\includegraphics[width=0.35\textwidth]{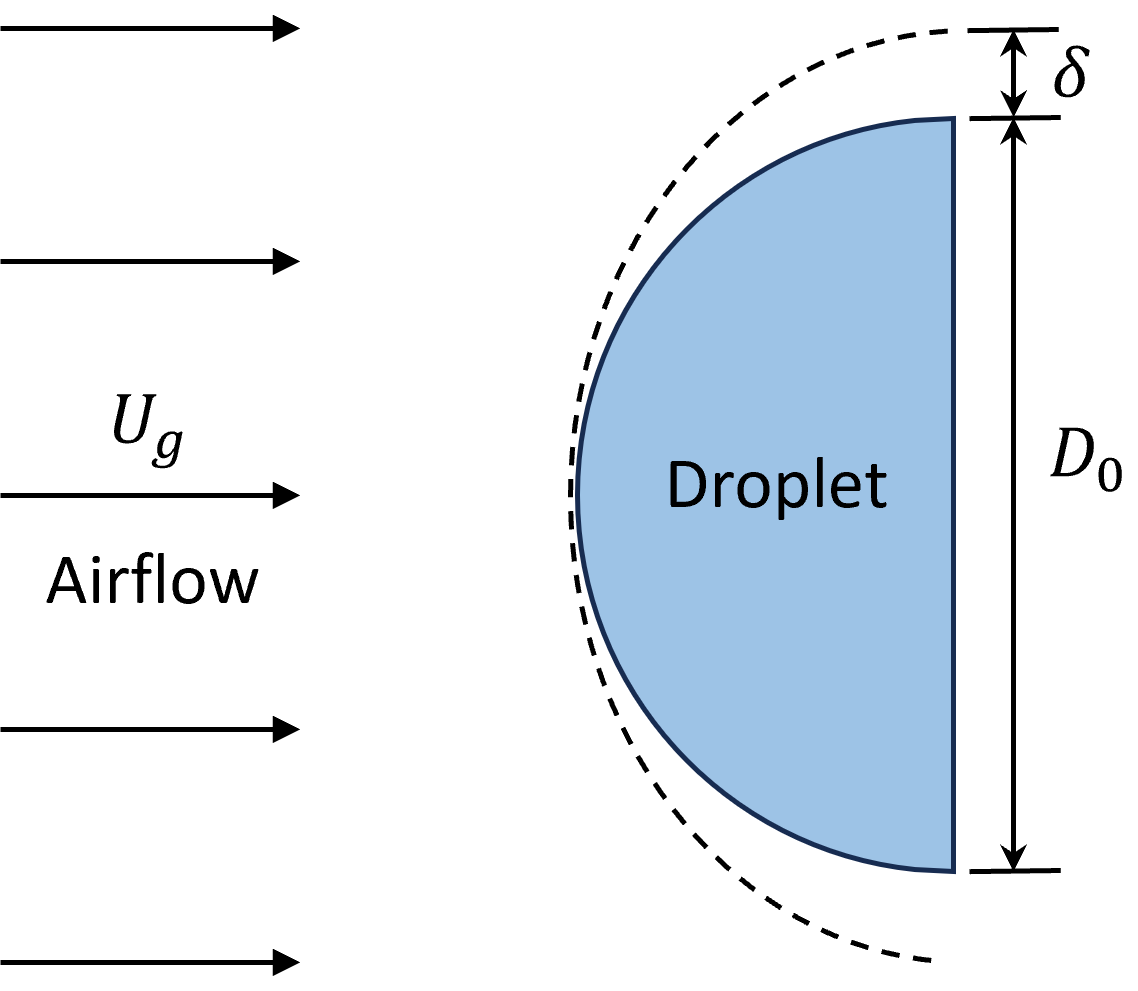}
\caption{Schematic representation of boundary layer (dashed curve) formed in the gas phase due to uniform airflow over a spherical droplet.}
\label{fig:boundary_layer}
\end{figure}

Concerning the relevant non-dimensional number, Weber number (Equation \ref{eq1}) represents the ratio of aerodynamic to the surface tension force, which is independent of liquid rheology, and it is equally important for Newtonian as well as viscoelastic droplets. The Ohnesorge number (Equation \ref{eq2}) accounts for the liquid viscosity, which is simple to calculate for the Newtonian and Boger fluids; however, it is not so straightforward for the shear-thinning fluids due to their shear-rate dependent viscosity variation. Theofanous et al. \cite{theofanous2013physics} proposed an effective Ohnesorge number $Oh_{eff}$, which accounts for the shear-thinning effect by considering an effective viscosity corresponding to the relevant shear rate in the liquid phase during droplet aerobreakup. To estimate the relevant shear rate, authors used the experimental observation from the transient evolution of the cross-stream diameter of the deforming droplet, which is actually an estimate of the elongation rate and not the shear rate. Although the internal flow in the liquid phase during aerobreakup is very difficult to analyze, an estimate of the shear rate, $\dot{\epsilon_s}$ can be obtained considering uniform gas flow over a spherical droplet with a boundary layer thickness, $\delta$ in the gas phase, as shown with the help of a schematic diagram in Figure \ref{fig:boundary_layer}. Shear stress continuity at the droplet periphery leads to the following expression-
\begin{equation}
\mu_g \frac{U_g}{\delta} \sim \mu_l \dot{\epsilon_s}
\label{eq3}
\end{equation}

Here, $\mu_g$ is the dynamic viscosity of the gas phase. Considering an order of magnitude estimate for aerobreakup scenario, $U_g\sim10^2$ m/s, $\mu_g\sim10^{-5}$ Pa-s, $\delta\sim10^{-6}$ m, minimum value of $\dot{\epsilon_s}$ comes out to be $\sim10^3$ s$^{-1}$ for the highest value of $\mu_l\sim10^0$ Pa-s (zero-shear viscosity of the ST2 fluid). Such a high value of $\dot{\epsilon_s}$ suggests that the infinite-shear viscosity of the shear-thinning fluid is a more suitable choice of parameter. However, it is important to note that in the case of polymeric liquids, the shear-thinning nature arises from the flow-induced re-arrangement of long polymer chains, which requires a certain amount of time \cite{bird1987dynamics}. This time is expected to be comparable to the relaxation time of the polymeric liquid, which for the present shear-thinning fluids is in the range of $\sim10^1-10^2$ ms. This timescale is much higher compared to the timescales of shock-induced droplet aerobreakup ($\sim10^2$ $\mu$s). Since the droplet aerobreakup happens at smaller timescales compared to the time required for the shear-thinning to come into play, it can be argued that the zero-shear viscosity is a more suitable choice of parameter for estimating the relevant non-dimensional numbers. Hence, we choose zero-shear viscosity for reporting the values of non-dimensional numbers for the case of shear-thinning fluids. Despite all the complexities discussed above, $Oh$ only accounts for the viscosity and not the elastic effects. Therefore, we use the elasticity number, $El$, which accounts for both the viscosity and elasticity of any viscoelastic liquid, defined as follows-
\begin{equation}
El = \frac{\lambda \mu_l}{\rho_l D_0^2}
\label{eq}
\end{equation}
$El$ and $Oh$ values for all the test liquids, considering an initial droplet diameter of 2 mm, are presented in Table \ref{table_1}.

\section{Results and Discussion} 
\label{result_discussion}
\begin{figure}[ht]%
\centering
\includegraphics[width=0.95\textwidth]{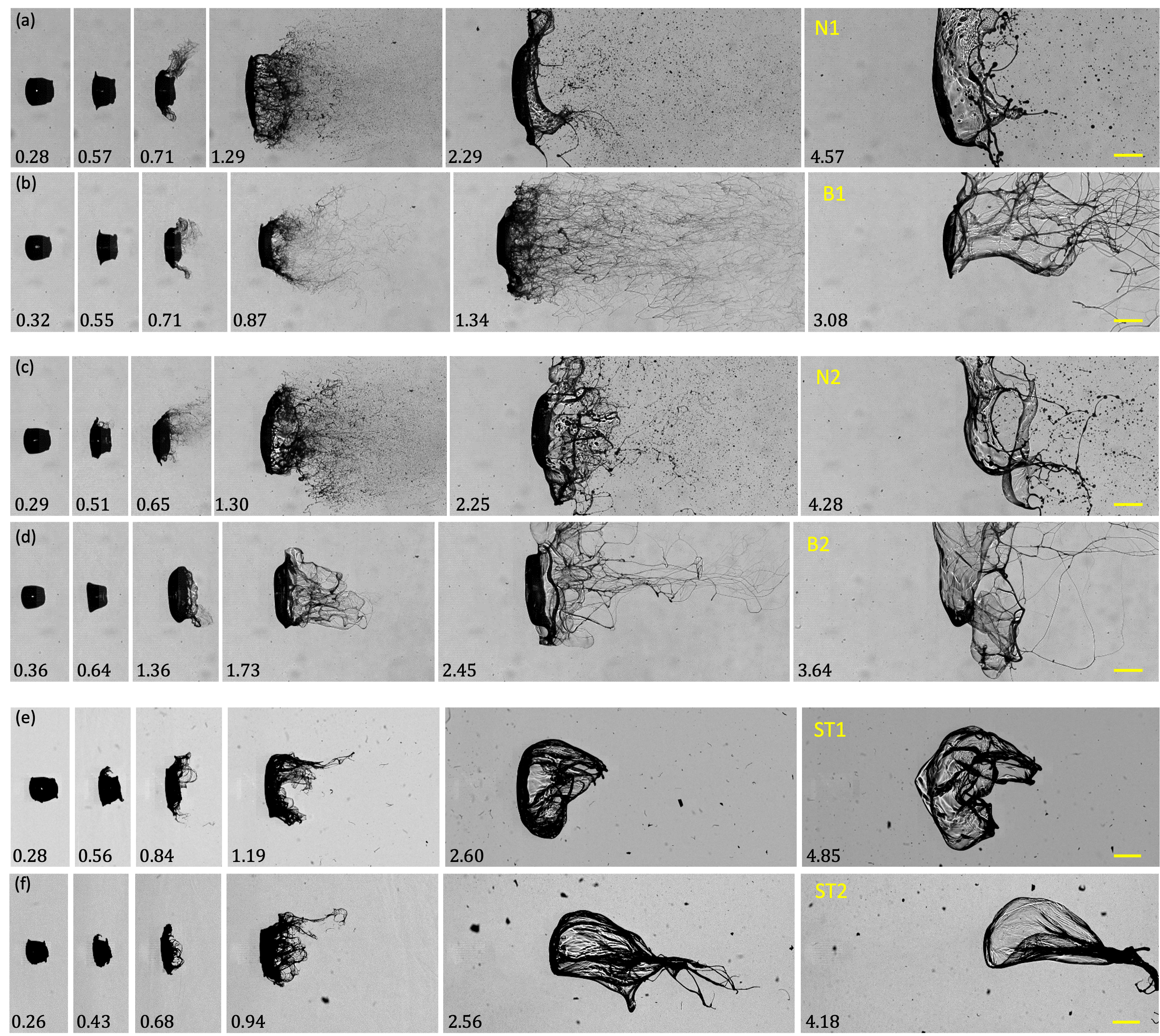}
\caption{Experimental images of droplet aerobreakup through SIE mode for all the test fluids at comparable Weber numbers. The fluid name and exact value of $We$ are- (a) N1, $We$=1422; (b) B1, $We$=1238; (c) N2, $We$=1307; (d) B2, $We$=1300; (e) ST1, $We$=1291; (f) ST2, $We$=1200. The experimental observations for ST1 and ST2 fluids are taken from our previous work \cite{chandra2023shock}. Airflow is from left to right, and the scalebar represents 2 mm. }
\label{fig:moderate_We}
\end{figure}
Figure \ref{fig:moderate_We} shows the experimental images of droplet aerobreakup at moderate $We$ ($\sim$1200-1400), where SIE mode of breakup is observed for all the test liquids. Non-dimensionalized time, $T=\frac{t}{\tau_I}$ is mentioned at the bottom of each image. Here, $t=0$ corresponds to the instant of droplet-shock wave interaction, and $\tau_I=\frac{D_0}{U_g} \sqrt{\frac{\rho_l}{\rho_g}}$ represent the inertial timescale, conventionally used as a suitable timescale in the description of droplet aerobreakup process \cite{nicholls1969aerodynamic}. To assess the role of liquid elasticity in the aerobreakup process, observations from the Newtonian droplet, N1 (Figure \ref{fig:moderate_We}a) can be compared with that of the viscoelastic Boger droplet, B1 (Figure \ref{fig:moderate_We}b), both of which have equal shear viscosity ($Oh\sim10^{-2}$). A similar comparison can be made for the N2 and B2 pair ($Oh\sim10^{-1}$), presented in Figure \ref{fig:moderate_We}c and \ref{fig:moderate_We}d, respectively. In all these cases, the characteristic of SIE mode can be observed at the early stages of breakup ($T<2$). This involves the deformation of the liquid droplet from spherical to a cupcake-like geometry and the formation of KH waves on the periphery of the droplet, leading to the ejection and subsequent entrainment of the liquid mass into the airflow. These early-stage breakup dynamics are similar for the Newtonian as well as the Boger fluid droplets, suggesting that liquid elasticity does not play a significant role in the early stages of aerobreakup. The major impact of liquid elasticity can be observed at later times ($T>2$ in Figure \ref{fig:moderate_We}) of droplet breakup. During this late stage, the liquid mass entrained in the airflow forms sheet and ligament structure, which, in the case of Newtonian liquids, undergoes further breakup to generate daughter droplets. Whereas, in the case of Boger fluid droplets, the presence of liquid elasticity provides significant resistance against the fragmentation of the liquid mass, and it remains an interconnected web of ligaments in the timescales of experimental observation. Similar observations are made for the aerobreakup of viscoelastic shear-thinning droplets (ST1 and ST2) as shown in Figure \ref{fig:moderate_We}e and \ref{fig:moderate_We}f. In these cases also, the early breakup characteristics of SIE mode are similar to the Newtonian droplet. Meanwhile, in the late stages, the fragmentation of the liquid sheet is resisted by the presence of elasticity, and hence, the final morphology is completely different. 

The impact of viscosity alone can be assessed by comparing the breakup of an N1 droplet (Figure \ref{fig:moderate_We}a), which is five times less viscous than that of the N2 droplet (\ref{fig:moderate_We}c). Considering the last images of these two panels, it can be observed that the ligaments emanating from the primary liquid core are slightly longer for the N2 case compared to the N1 case. While it is apparent that an increase in viscosity contributes to a certain delay in the breakup process, this delay is not substantial enough to completely hinder fragmentation within the experimental timescales. Consequently, both N1 and N2 droplets exhibit the formation of daughter droplets through ligament breakup. In contrast, the persistence of stable sheets and ligaments in the case of viscoelastic droplets (Boger as well as shear-thinning) shows that the resistance to fragmentation offered by elasticity is manifold higher than that of the viscosity. This is also supported by the static capillary breakup experiments performed for DoS rheometry (Figure \ref{fig:rheology}c), where the ligament breakup slows down dramatically when the elastic effects become dominant. This emphasizes that the final morphology of the liquid mass in the aerobreakup of a viscoelastic droplet is governed primarily by the elasticity compared to other rheological properties.

\begin{figure}[ht]%
\centering
\includegraphics[width=0.8\textwidth]{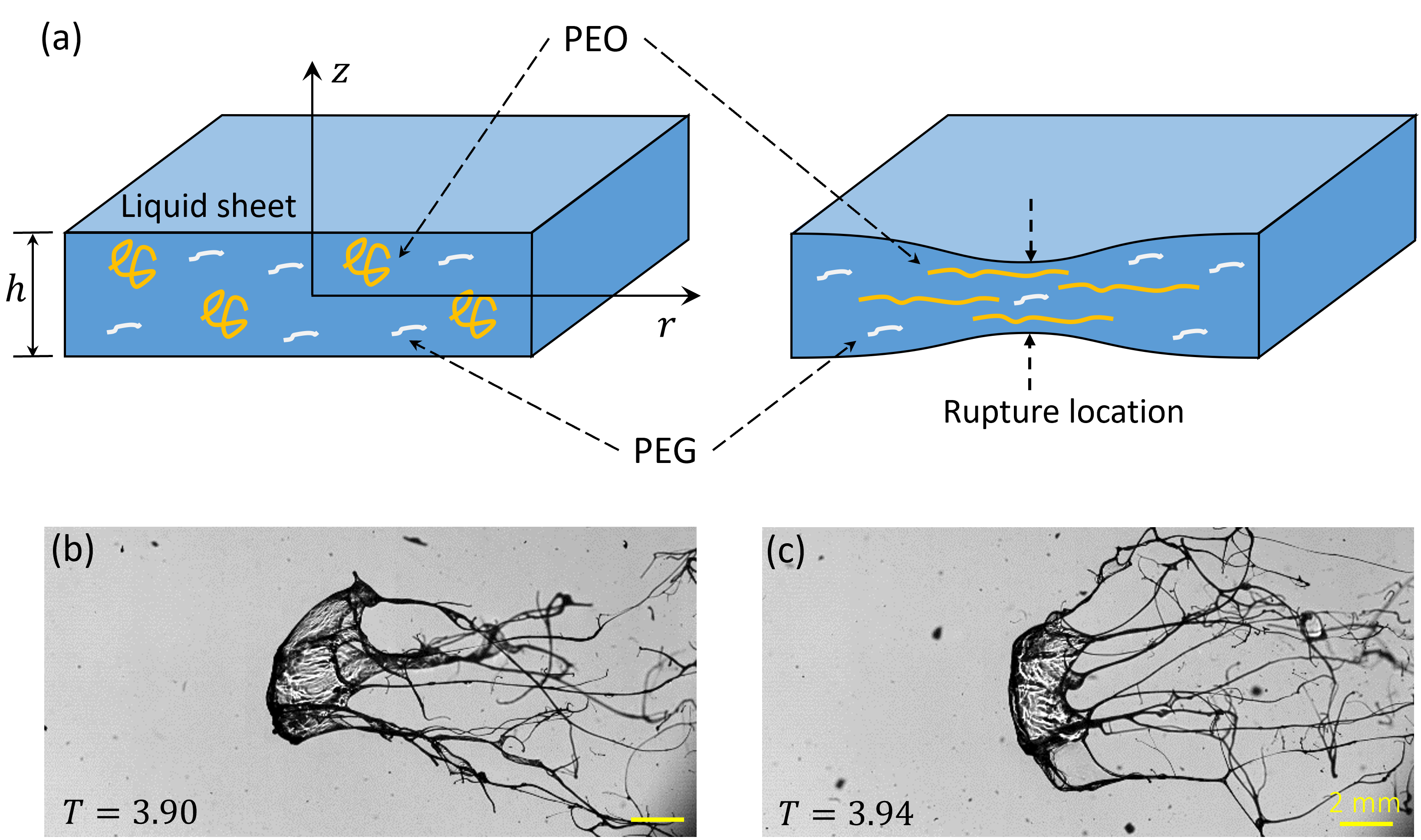}
\caption{(a) Schematic showing the extension of long chain polymer molecule during the liquid sheet breakup. Liquid mass morphology observed in the late stage of aerobreakup for aqueous solutions of PEO with concentrations (b) 0.1\% and (c) 0.042\%.}
\label{fig:polymer-stretch}
\end{figure}

Our previous work \cite{chandra2023shock} showed that a well-defined hierarchy of liquid mass morphology can be observed in the late stage of SIE breakup mode depending on the elasticity number. These morphology with their decreasing order of $El$ are- stable sheet structure (SS), sheet breakup-stable ligament (SB-SL), bead-on-a-string (BOAS) structure, and sheet breakup-ligament breakup (SB-LB). However, in the present work, we note an anomalous trend when comparing the SIE mode breakup results of ST1 (Figure \ref{fig:moderate_We}e) and B2 (Figure \ref{fig:moderate_We}d) droplet. SIE breakup of an ST1 droplet results in a stable sheet (SS) structure despite having lower $El$ ($\approx$ 0.9) compared to the B2 droplet ($El\approx$ 4.2) which gives sheet breakup-stable ligament (SB-SL) structure. This suggests that $El$ is not a universal non-dimensional number to categorize the breakup morphology obtained from the aerobreakup of a viscoelastic droplet. One of the possible explanations for this anomaly may come from looking at the concentration of long-chain polymer molecules and their contribution to inhibiting the liquid sheet breakup. Schematic shown in Figure \ref{fig:polymer-stretch}a depicts the cross-section of a polymeric liquid sheet with an initial uniform thickness, $h$ in the $z-$direction and its spread in the $r-$direction. Although the exact cause and mechanism of liquid sheet breakup is not yet known \cite{jackiw2022prediction, neel2018spontaneous}, the sheet must experience a local thickness reduction near the rupture location, leading to a bi-axial extensional flow in the liquid phase. In such extensional flows, a long-chain polymer molecule undergoes a transition from a coil to stretched conformation \cite{rajesh2022transition}, resulting in a large elastic stress and hence resisting the sheet breakup (Figure \ref{fig:polymer-stretch}a). Boger fluids (B1, B2) used in the present work contain two different types of polymers, a long chain polymer PEO with molecular weight $5\times10^6$ g/mol and a comparatively smaller chain polymer PEG with molecular weight $10^4$ g/mol. The PEG molecules are sufficiently small ($\sim5\times10^2$ times smaller than the fully stretched length of a PEO molecule) so that they do not contribute towards the elasticity of the Boger fluids \cite{dontula1998model}. The main contribution to liquid elasticity comes from the stretching of long-chain PEO molecules. Since the PEO concentration in ST1 fluid is higher (0.4\%) compared to the B2 fluid (0.15\%), it can be explained that ST1 fluid results in stable sheet structure while the B2 fluid results in sheet breakup-stable ligament morphology. This hypothesis is further supported by our previous experiments performed using two different polymeric solutions containing only PEO (not PEG) with concentrations of 0.1\% ($El\approx3\times10^{-3}$) and 0.042\% ($El\approx1\times10^{-3}$) \cite{chandra2023shock}. These fluids have approximately the same PEO concentration as B2 (0.15\%) and B1 (0.05\%) fluid, and they also result in sheet breakup-stable ligament morphology (Figure \ref{fig:polymer-stretch}b and \ref{fig:polymer-stretch}c) during the late stage of SIE breakup mode similar to B2 and B1 fluid. This favors the argument that the concentration of long-chain polymer, which contributes towards the extensional elasticity, controls the final breakup morphology. Despite the above discussion, it is not guaranteed that the concentration of the long-chain polymer molecule is the universal parameter deciding the breakup morphology because we have explored only one type of PEO (molecular weight $=5\times10^6$ g/mol). Further research with more experimental data and theoretical analysis is required to confirm such universal parameters.

\begin{figure}[ht]%
\centering
\includegraphics[width=0.95\textwidth]{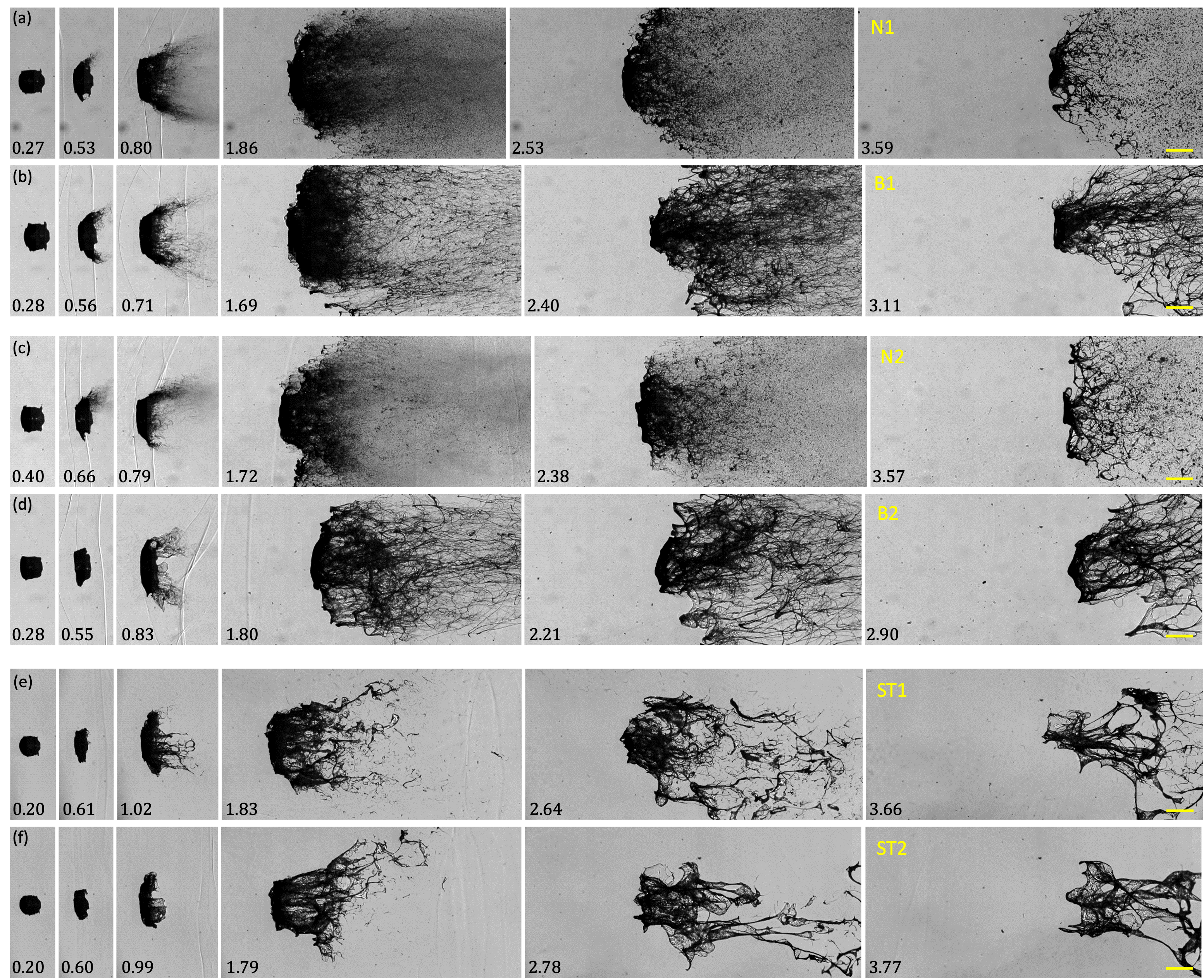}
\caption{Experimental images of droplet aerobreakup at high Weber numbers where catastrophic mode is observed. The fluid name and exact value of $We$ are- (a) N1, $We$=3869; (b) B1, $We$=3724; (c) N2, $We$=4076; (d) B2, $We$=3082; (e) ST1, $We$=5362; (f) ST2, $We$=4898. The experimental observations for ST1 and ST2 fluids are taken from our previous work \cite{chandra2023shock}. Airflow is from left to right, and the scalebar represents 2 mm. }
\label{fig:high_we}
\end{figure}

Figure \ref{fig:high_we} shows the experimental images of droplet aerobreakup at sufficiently high $We$ ($>\approx$ 3000) where the catastrophic breakup mode is observed for all the test liquids. The catastrophic breakup mode is assisted by the RT instability waves, in addition to the droplet deformation and shear-induced entrainment of liquid mass observed in SIE mode \cite{chandra2023shock, joseph2002rayleigh, mansoor2023investigation}. The RT waves, appearing as corrugations on the windward face, facilitate the penetration of airflow into the bulk liquid mass, leading to its widespread catastrophic breakup. From Figure \ref{fig:high_we}, it can be observed that the early-stage ($T\leq2$) breakup characteristics of catastrophic mode for Boger fluid droplets (B1, B2) are similar to their Newtonian counterparts (N1, N2) as well as the shear-thinning droplets (ST1, ST2). However, the final morphology of the liquid mass is altered by the presence of liquid elasticity.

From the observations and related discussion provided in Figure \ref{fig:moderate_We} (SIE mode) and Figure \ref{fig:high_we} (catastrophic mode), it is clear that the liquid elasticity plays a negligible role during the early stages of aerobreakup and hence the breakup modes are same as their Newtonian counterpart. The dominant effect of liquid elasticity appears during the late stages of aerobreakup in terms of the morphology of the fragmenting liquid mass. We had the same conclusion from our previous work \cite{chandra2023shock} but with shear-thinning viscoelastic droplets. The present work confirms our previous claim on the role of liquid elasticity in modulating aerobreakup. However, now the confirmation comes from Boger fluid droplets, which ensures that any deviation in the final morphology compared to the Newtonian droplets is only due to elasticity and not due to the shear-thinning behavior.

\subsection{Regime plot of droplet aerobreakup}\label{regime}
\begin{figure}[ht]%
\centering
\includegraphics[width=0.8\textwidth]{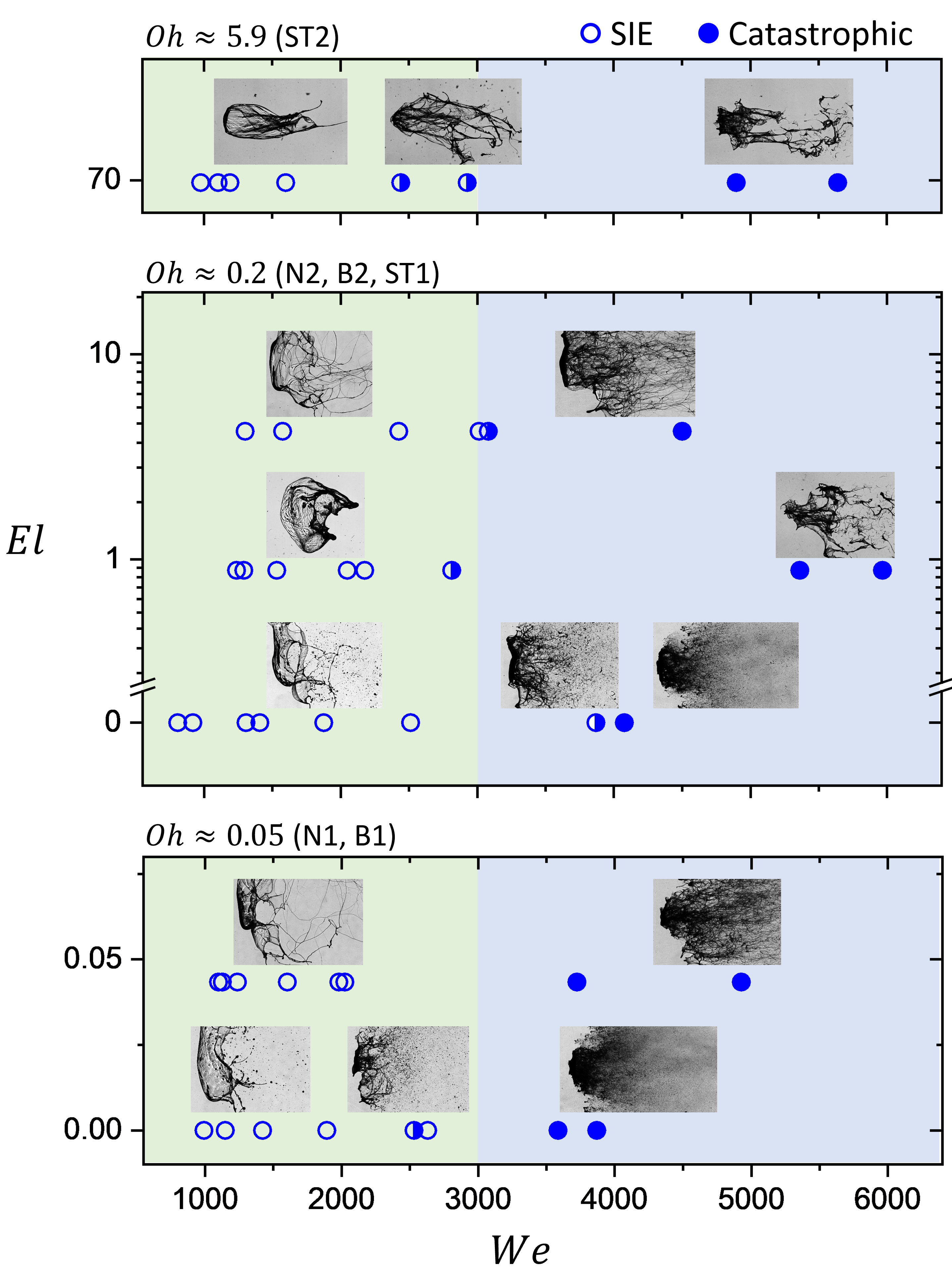}
\caption{Regime plot of the droplet breakup mode with inset as the experimental images showing breakup morphology. The open symbol represents the SIE breakup mode, the filled symbol represents the catastrophic mode, and the partially filled symbol represents the transition between the two regimes. The data for the ST1 and ST2 droplets are taken from our previous work \cite{chandra2023shock}.}
\label{fig:regime_plot}
\end{figure}
The data points from the present work (N1, N2, B1, B2) and our previous work (ST1, ST2) \cite{chandra2023shock} are collated in the form of a regime plot in Figure \ref{fig:regime_plot}. Inset in this regime plot shows the representative experimental images of the breakup morphology obtained during the late stage of aerobreakup. The abscissa and the ordinate axes in the regime plot correspond to $We$ and $El$, respectively. It should be noted that these two axes are not sufficient to identify all the data points uniquely. For instance, N1 and N2 fluids both correspond to $El=0$ while they have different shear viscosity and hence different $Oh$. Similarly, the N1-B1 and N2-B2 pairs have the same $Oh$ but different $El$. Therefore, a third axis corresponding to $Oh$ is required to properly present all the data points. Since the $Oh$ axis has limited data points in the present work, they are shown as different stacks in Figure \ref{fig:regime_plot}. $Oh\approx0.05$ corresponds to the N1 and B1 fluids, $Oh\approx0.2$ corresponds to the N2, B2 and ST1 fluids, and $Oh\approx5.9$ presents the ST2 fluid. The regimes of SIE and catastrophic mode are shaded as green and blue, respectively. It can be observed that for the present range of parameters, the breakup modes are independent of $El$ and $Oh$, and they are decided only by the $We$ value. For representation purposes, $We=3000$ is shown as the boundary between SIE and catastrophic mode. However, it should be noted that the breakup modes are decided by the visual inspection of experimental observation, and in some cases, it becomes subjective to categorize them near the boundary ($We\approx3000\pm500$) which is like a transition between the two regimes. Experimental images provided in Figure \ref{fig:regime_plot} makes it clear that the liquid elasticity significantly alters the breakup morphology by resisting the fragmentation of liquid sheets and ligaments. This provides a direction for future research in the area of viscoelastic droplet breakup. Unlike the Newtonian droplet aerobreakup, where the regime plot on a $We-Oh$ plane is based upon the initial morphology of breakup (bag, bag-stamen, multi-bag, shear stripping, catastrophic) \cite{HSIANG_1992}, the study related to viscoelastic droplets should aim for identifying the final morphology of the breakup.

\section{Conclusion}\label{conclusion}
The present experimental study provides insights into the influence of liquid elasticity in the aerobreakup of a polymeric droplet. Experiments cover a Weber number range of $\sim$ 800 to 5000, where shear-induced entrainment (SIE) and catastrophic breakup modes are observed. Boger fluids (B1, B2) with matched shear viscosity to their Newtonian counterparts (N1, N2) are carefully selected to isolate the effect of liquid elasticity from shear-thinning behavior. The findings indicate that during the early stages of aerobreakup, liquid elasticity has a negligible impact, resulting in breakup modes similar to Newtonian droplets for a given Weber number. However, in the late stages of aerobreakup, liquid elasticity emerges as a significant factor, offering substantial resistance to the fragmentation of the liquid sheet and ligaments. Consequently, this leads to a different morphology of the liquid mass compared to Newtonian droplets. This observation aligns with the conclusions drawn in our previous work \cite{chandra2023shock}, which involved shear-thinning polymeric droplets. Thus, the current study affirms that the identified influence of liquid elasticity persists even in the absence of shear-thinning behavior. Comparing the SIE mode breakup morphology of a B2 droplet ($El=4.21$) with that of an ST1 droplet ($El=0.89$) revealed that despite having smaller $El$, the ST1 droplet results in a more stable sheet than B2. This suggests that $El$ is not a universal parameter in deciding breakup morphology. An explanation for this could be that only the long-chain polymer (PEO in the present case) contributing to the extensional elasticity of polymeric liquid provides significant resistance against breakup. Since the PEO concentration is higher in ST1 than in B2, a stable sheet structure is observed for the case of ST1 droplet.

\backmatter

\bmhead{Author contributions}
NKC contributed to data acquisition, investigation, software, writing original draft; SS contributed to data acquisition, investigation, writing original draft; SB contributed to funding acquisition, project administration, supervision, writing original draft; AK contributed to funding acquisition, project administration, supervision, conceptualization, writing original draft.

\section*{Declarations}\label{Declarations}

\bmhead{Ethical Approval}
Not applicable.

\bmhead{Funding}
AK acknowledges partial funding from the Science and Engineering Research Board (SERB) through grant number CRG/2022/005381. NKC acknowledges support from the Prime Minister's Research Fellowship (PMRF).

\bmhead{Availability of data and materials}
The data that support the findings of this study are available from the corresponding author upon reasonable request.

%%\bibliography{sn-bibliography}% common bib file
%% if required, the content of .bbl file can be included here once bbl is generated
%%\input sn-article.bbl

\bibliography{sn-bibliography}
\bibliographystyle{vancouver}

\end{document}